# Hidden structural order controls Li-ion transport in cation-disordered oxides for rechargeable lithium batteries

Huiwen Ji[†,‡], Alexander Urban[†,‡], Daniil A. Kitchaev[¶], Deok-Hwang Kwon[†,‡], Nongnuch Artrith[†,‡], Colin Ophus[&], Wenxuan Huang[¶], Zijian Cai[†], Tan Shi[†,‡], Jae Chul Kim[‡], Gerbrand Ceder[†,‡,*]

[†]*Department of Materials Science and Engineering, University of California Berkeley, Berkeley, CA 94720, USA*

[‡]*Materials Sciences Division, Lawrence Berkeley National Laboratory, Berkeley, CA 94720, USA*

[¶]*Department of Materials Science and Engineering, Massachusetts Institute of Technology, Cambridge, MA 02139, USA*

[&]*National Center for Electron Microscopy, Molecular Foundry, Lawrence Berkeley National Laboratory, Berkeley, CA 94720, USA*

*Email: gceder@berkeley.edu

# Abstract

Crystal structures play a vital role in determining materials properties. In Li-ion cathodes, the crystal structure defines the dimensionality and connectivity of interstitial sites, thus determining Li-ion diffusion kinetics. While a perfect crystal has infinite structural coherence, a class of recently discovered high-capacity cathodes, Li-excess cation-disordered rocksalts, falls on the other end of the spectrum: Their cation sublattices are assumed to be randomly populated by Li and transition metal ions with zero configurational coherence based on conventional X-ray diffraction, such that the Li transport is purely determined by statistical effects. In contrast to this prevailing view, we reveal that cation short-range order, hidden in diffraction, is ubiquitous in these long-range disordered materials and controls the local and macroscopic environments for Li-ion transport. Our work not only discovers a crucial property that has previously been overlooked, but also provides new guidelines for designing and engineering disordered rocksalts cathode materials.

**Introduction**

The development of cost-effective Li-ion batteries depends on the discovery of high-energy-density cathode materials composed of nonprecious elements.[1] Rational design of cathodes requires an understanding of the precise role that each chemical component has in determining performance. Traditionally, one thinks of redox active elements, such as Ni and Co, and stabilizers, such as $Mn^{4+}$ in NMC-class materials.[2] We demonstrate in this paper that in the recently discovered class of Li-excess cation-disordered rocksalt cathodes (DRX), the chemistry of then non-redox active stabilizer plays a critical role in performance through subtle structural changes.

Disordered rocksalt materials were recently shown to have facile Li-transport enabled by a percolating network of Li-rich environments[3]. Their ability to function without requiring cation ordering has enabled novel cathodes with remarkable chemical diversity[3, 4]. Many new cathode materials, in some cases containing only earth-abundant elements (e.g., Fe, Mn, and Ti) have been developed in this category, such as $Li_{1.3}Mn_{0.4}Nb_{0.3}O_2$[5], $Li_{1.2}Mn_{0.4}Ti_{0.4}O_2$[6], $Li_{1.2}Ni_{1/3}Ti_{1/3}Mo_{2/15}O_2$[7], $Li_4Mn_2O_5$[8], and $Li_2FeV_{0.5}Ti_{0.5}O_4$[9] as well as their fluorinated variants[10-16]. A prevailing assumption when studying DRX cathodes is that all the cation species are randomly distributed. However, we show in this paper that even minor deviations from randomness, not detectable by typical X-ray diffraction (XRD), can have profound influence on performance.

Figure 1a presents a typical DRX crystal structure. The highlighted tetrahedron represents a channel through which Li migrates[17] (Figure 1b). The Li migration barrier depends on the tetrahedron height [3, 18-20] and the number of transition metal (TM) ions within the environment (i.e. 0-TM, 1-TM, or 2-TM).[4] If the cation arrangement is random, any DRX with the same Li to TM ratio should have an equivalent distribution of Li-migration channel types and thus similar Li transport properties.

In this study, we compare two Li-excess DRXs, $Li_{1.2}Mn_{0.4}Ti_{0.4}O_2$ (LMTO) and $Li_{1.2}Mn_{0.4}Zr_{0.4}O_2$ (LMZO). Based on their chemical similarity, these materials would be expected to have comparable electrochemical properties, as $Zr^{4+}$ and $Ti^{4+}$ are isoelectronic and their sole role is to charge compensate for the excess Li. If anything, the larger ionic radius of $Zr^{4+}$ should result in a larger lattice parameter for LMZO, which is generally considered beneficial for Li mobility.[18-20] However, contrary to these expectations, we observe that the performance of LMTO is considerably better than that of LMZO. We reveal through a combination of electron diffraction, neutron pair distribution function measurements, and cluster-expansion Monte Carlo simulation that the difference in the performance of LMTO and LMZO is due to different cation short-range order (SRO), which controls the population and connectivity of Li-migration channels. We further identify general rules that govern the relationship between SRO and Li transport by expanding our analysis to other combinations of TMs. These results indicate the importance of SRO and provide another important handle to tailor the performance of DRX cathode materials, in addition to the already large compositional flexibility.

## Results

**Synthesis and electrochemistry**

We synthesized LMTO and LMZO using a solid-state method and verified the compositions of the products to be nearly identical to the target compositions using elemental analysis (Table S1). XRD patterns (Figure 1c) reveal a DRX structure. Rietveld refinement indicates that the lattice parameter of LMZO ($a$ = 4.27 Å) is larger than that of LMTO ($a$ = 4.15 Å), as expected. Scanning electron microscopy (SEM) images of shaker-milled LMTO and LMZO (Figure 1d and 1e) confirm particle sizes of ~100 nm for both materials. The galvanostatic voltage profiles of LMZO and LMTO are presented in Figure 1f and 1g. Consistent with a previous report by Yabuuchi *et al.*,[6] LMTO delivers a large first-cycle capacity of approximately 260 mAh/g at room temperature, which corresponds to 0.79 Li/*f.u.* However, LMZO delivers a much smaller discharge capacity (0.52 Li/*f.u.*). In addition, the average discharge voltage for the first cycle is lower in LMZO (2.75 V) than in LMTO (3.08 V).

The Li diffusion is a key factor in determining the observed capacity[21, 22]. To test whether the capacity of LMZO is limited by kinetics, galvanostatic cycling at both room-temperature and 50 °C was performed, as shown in Figure 2a and 2b. The difference between room-temperature and 50 °C indicates pronounced kinetic limitations in LMZO over the entire voltage range, unlike LMTO, which only shows improvement at 50 °C near the upper cutoff voltage. As shown in Figure 2c, at 50 °C, LMZO delivers a reversible capacity of 0.80 Li/*f.u.*, a 54% increase from that at room temperature. In contrast, the capacity of LMTO improves by only 27% to 1.0 Li/*f.u.*, when cycled at 50 °C. The Li chemical diffusivities ($D_{Li}$) of LMTO and LMZO were determined using the potentiostatic intermittent titration technique (PITT)[23-25] during the initial charge from open-circuit voltages to 4.7 V (Figure 2d). The chemical diffusivities of LMTO and LMZO have distinct regions that slightly differ for the two materials; nevertheless, the Li diffusivity in LMTO is much higher than that in LMZO, confirming that the capacity of LMZO is limited by Li transport kinetics.

**Local-structure characterization**

The discovery that two almost identical materials exhibit significantly different Li transport led us to investigate the subtle structural differences between LMTO and LMZO. Figure 3a and 3c present the electron diffraction (ED) patterns of LMTO and LMZO, respectively. Aside from the reflection spots that can be indexed with a DRX structure, we also observe diffuse scattering patterns surrounding the Bragg reflections, suggesting the existence of SRO. The formation of SRO can be understood as a preferred local arrangement of species, resulting in non-vanishing patterns in reciprocal space.[26-29] Notably, the diffuse scattering patterns are completely different, both in shape and orientation, for the two materials, indicating significant difference in SRO. Based on previous characterization of SRO in oxides and alloys, the SRO patterns in LMTO is characteristic of octahedral cation clusters similar to the [$Li_3Fe_3$] in cubic-$LiFeO_2$[30], whereas that in LMZO is likely associated with tetrahedral cation clusters[31]. In addition, the intensity of the diffuse scattering pattern of LMZO is noticeably stronger at several maxima highlighted with yellow arrows, suggesting more pronounced SRO in LMZO than in LMTO. These characteristic features observed in the diffuse scattering patterns are well reproduced by simulation (Figure 3b and 3d) based on thermodynamically representative structures, which we obtain from Monte Carlo (MC) sampling at 1000°C with a cluster expansion (CE) Hamiltonian parameterized to fit the rock-salt configurational energies derived from density-functional theory (DFT).[32, 33] Detailed analysis of the model structures is presented in the next section.

Neutron pair distribution function (NPDF) measurements were performed to precisely characterize the SRO in LMTO and LMZO. In NPDF analysis, Fourier transformation of the total scattering data to real space is performed, thereby providing additional information about SRO that is hidden in diffraction patterns[34]. The refinement of the NPDF patterns of LMTO and LMZO is presented in Figure 3e–h. We use two structural models for the refinement: One is a random model that assumes a random cation distribution in a distortion-free lattice, and the other is the MC-derived structural model described above. The random model produces a reasonable fit for LMTO but not for LMZO, as indicated by the goodness-of-fit values ($R_w$). In LMZO, the simulation based on the random model differs significantly from the experimental observation near 2, 3, 5, and 8 Å (Figure 3g). Nevertheless, in the longer range, the random model produces a good fit for both materials (Figure S6). These results suggest that LMZO has significantly more SRO than LMTO. As a comparison, the refinement using the MC configurations, presented in Figure 3f and 3h, shows significant improvement for both compounds.

Combining the analysis of ED and NPDF, we find that LMTO and LMZO differ in their SRO and that the ab initio MC structures simulated near the synthesis temperature are precise manifestations of the SRO in these materials.

**Computational modeling of Li transport environments in LMTO and LMZO**

The structures obtained from MC simulation uncover the atomistic nature of the SRO, enabling further analysis of the local and macroscopic Li-transport environments in LMTO and LMZO.

In DRX materials, local environments can be characterized by the occurrence of cation clusters, among which the ones most relevant to Li transport are tetrahedral clusters, i.e., Li migration channels[21]. Because of their connectivity in the structure, their population is not completely independent, a phenomenon that has been recognized early on when studying the entropy of FCC systems [35]. Figure 4a summarizes the occurrence of various tetrahedral clusters in LMTO and LMZO relative to a random case. We observe that the occurrence of $Li_4$ tetrahedra (i.e. 0-TM channels), which is the most important for good Li transport, is significantly lower in LMZO than in LMTO, although both materials have lower $Li_4$ population than for a random cation distribution. Conversely, the population of $Li_3M$ tetrahedra, i.e. 1-TM channels, is much higher in LMZO than in LMTO. More specifically, the $Li_3Zr$ clusters account for 31% of all cation tetrahedra in LMZO, as compared to 22% for $Li_3Ti$ in LMTO and 17% for $Li_3M'$ in the random case (Table S5). Such a high population of 1-TM channels in LMZO is detrimental for Li transport as it was previously demonstrated that in a typical DRX, the migration barrier through a 1-TM channel is on average 200 meV higher than that through a 0-TM channel.[3] These observations indicate that SRO strongly modifies the population of local tetrahedral clusters: LMTO favors $Li_4$ clusters more than LMZO does; while LMZO contains more $Li_3M$ (especially $Li_3Zr$) clusters.

While the population of $Li_4$ tetrahedra is critical for local Li migration, a sufficient connectivity between these environments is another key criterion to ensure macroscopic Li transport. Figure 4b presents a connectivity analysis for LMTO and LMZO as compared to a random cation distribution. A connectivity plot, averaged over 600 sampled MC structures, shows the fraction of Li content in Li networks of more than a certain number of 0-TM inter-connected Li sites. The fraction of Li content initially decreases with an increasing network size and finally plateaus at the percolating Li level, which is considered the lower-bound of accessible Li. Figure 4b suggests that LMTO has more extensively connected 0-TM Li networks compared to LMZO, although both materials show worse Li connectivity than the random case. Specifically, in LMTO, nearly 40% of Li is in 0-TM networks of more than 100 Li sites and approximately 35% of Li is percolating. In contrast, in LMZO, the fraction of Li in networks of more than 25 Li sites is already vanishingly small and the material is not percolating.

To visualize the Li diffusion pathways in LMTO and LMZO, representative MC structures are shown in Figure 4c. The 0-TM connected Li networks are highlighted in green. From these images, it is clear that LMTO has extensive 0-TM Li networks that are well connected and should allow facile Li transport, whereas LMZO lacks 0-TM channels, thereby impeding Li diffusion.

**Chemistry dependence of SRO and Li transport environments**

With the understanding of how SRO affects Li transport in LMZO and LMTO, we can investigate other combinations of TMs to determine how the SRO-affected Li transport environments vary with chemistry. Figure 5 shows the accessible Li contents based on the percolation theory[3] in a variety of $Li_{1.2}M'_aM''_bO_2$ compounds under two conditions: (i) allowing only 0-TM jumps or (ii) allowing any given Li to make a single 1-TM jump before reaching the 0-TM percolating network. The rationale behind the chosen conditions is that although sufficient 0-TM channels are required for macroscopic Li transport, in reality, on the atomic scale Li ions can occasionally overcome higher migration barriers through 1-TM channels on the time scale of battery charge and discharge.

Consistent with the connectivity analysis, LMTO has a high accessible Li content of 35% with only 0-TM jumps, which increases to nearly 58% by allowing 1-TM jumps, whereas LMZO is not percolating under either condition. The accessible Li contents for $Mn^{3+}$–$Nb^{5+}$ are worse than LMTO but still significantly better than those of LMZO, consistent with the good performance of the $Mn^{3+}$–$Nb^{5+}$ materials at elevated temperatures.[5] $V^{3+}$–$Nb^{5+}$ is quite similar to $Mn^{3+}$–$Nb^{5+}$, which might explain the limited first-cycle capacity of a previously reported Li-V-Nb-O DRX, which continuously increases upon cycling as the structure gets more disordered with V migration[36]. Overall, in all cases, compounds containing $Ti^{4+}$ and $Mo^{6+}$ lead to higher accessible Li contents than those containing $Nb^{5+}$. Moving to divalent TM ions such as $Ni^{2+}$ and $Co^{2+}$, we find that they generally have higher accessible Li contents than the trivalent analogues, except that $Mn^{2+}$–$Nb^{5+}$ appears to be a poor-performing exception.

**Discussion**

We have shown that SRO is critical in controlling Li-conductive environments and has a general dependence on chemistry. The remaining questions are thus what the microscopic origin of these trends is and how we can predict and manipulate SRO for the benefit of Li transport. Although the use of MC simulation is necessary to precisely reproduce SRO, empirical rules can be derived for intuitive prediction.

We find that the charge and size effects, which determine the stability of solid-state materials, also explain the trends in SRO. On the one hand, the high-valent TMs (e.g., $Mn^{3+}$, $Ti^{4+}$, $Nb^{5+}$) in DRXs tend to repel each other and intimately mix with $Li^+$ in order to keep local electroneutrality, thereby inhibiting Li segregation into $Li_4$ tetrahedra. This charge effect becomes more pronounced with increasing metal valence. On the other hand, the size mismatch between high-valent TMs and $Li^+$ facilitates Li segregation in order to minimize strain, counteracting the charge effect. The competition between the two effects is best demonstrated in the case of LMTO and LMZO, where $Zr^{4+}$ exhibits a stronger net attraction to $Li^+$ than $Ti^{4+}$ despite their common valence (Figure S8). One explanation for this phenomenon is that $Ti^{4+}$ (0.605 Å) is much smaller than $Li^+$ (0.76 Å), and therefore, the size effect tends to segregate $Li^+$ from $Ti^{4+}$. In contrast, the size of $Zr^{4+}$ (0.72 Å) is close to that of $Li^+$, meaning that electrostatics dominates the size effect, favoring a maximal separation between the high-valent $Zr^{4+}$ and a corresponding local ordering between $Zr^{4+}$ and $Li^+$.

However, the prediction of SRO becomes elusive when comparing DRXs with non-isoelectronic TMs, e.g., $Li_{1.2}Mn_{0.4}Zr_{0.4}O_2$ and $Li_{1.2}Mn_{0.6}Nb_{0.2}O_2$. In this case, there is a more complex tradeoff between interaction strengths and TM concentration in determining the degree of local ordering and Li segregation. Figure 6a shows a more intuitive relationship between accessible Li and chemistry, for DRXs composed of trivalent redox-active TMs. We observe a consistent negative correlation between the accessible Li content and the average TM ionic radius. The rationale behind the chosen $x$ axis is that in these DRXs, with a fixed Li-excess level, the average valence of the TMs is fixed accordingly and so is the electrostatic interaction strength between Li and TMs. Consequently, the size effect becomes dominant. Based on this correlation, we can further predict that high-valent metal species with large ionic sizes, e.g., Sc (0.745 Å) and In (0.80 Å), are likely to mix with Li and impede Li diffusion, whereas others with small sizes, e.g., $Ga^{3+}$ (0.62 Å) and $Sb^{5+}$ (0.60 Å), are likely to facilitate local Li segregation and efficient transport.

DRXs containing divalent TMs form another unique class because divalent ions have the proper average metal valence in a DRX oxide and therefore do not require mixing with $Li^+$ to keep local electroneutrality. Additionally, they do not repel high-valent metals as much as trivalent ions do (Figure S8) but can in turn mix with the high-valent metals to buffer their attraction to Li. This buffer effect explains why DRXs containing divalent TMs generally have better $Li_4$ segregation (Figure S9) and higher accessible Li contents (Figure 5) than their trivalent analogues. Figure 6b shows the accessible Li contents of DRXs containing divalent redox-active TMs and various stabilizers. We find that compounds containing $Ti^{4+}$ and $Mo^{6+}$ generally have higher accessible Li contents than the ones containing $Nb^{5+}$, a phenomenon possibly explained by the tradeoff between the electrostatic interaction strength and metal concentration. In addition, within each plot, we observe a consistent dependence on the divalent-TM radius. A feasible mechanism would be that as the ionic radius of the divalent TM increases, the increasing size mismatch tends to segregate the divalent TMs from the high-valent TMs, weakening the buffer effect, which eventually vanishes for $Mn^{2+}$ because its ionic radius is even larger than that of $Li^+$.

Overall, we find that SRO controls the Li transport in DRXs by altering the distribution of 0-TM, 1-TM, and 2-TM channels as well as the connectivity between them. This observation is in contrast to stoichiometric layered oxides (e.g., $LiNi_{0.5}Mn_{0.5}O_2$[37] and $Li[Ni_xMn_xCo_{1-2x}]O_2$[38, 39]) where Li and TMs are well separated and any SRO in the TM layer imposes minimal impact on overall Li transport kinetics. We have also identified a few guidelines for the manipulation of SRO. (i) For DRXs containing high-valent metals, the average TM ionic radius is an important metric to measure the degree of Li segregation. The use of large metal ions such as $Zr^{4+}$, $Sc^{3+}$, and $In^{3+}$ should be minimized. (ii) DRXs containing divalent TMs often facilitate Li segregation compared to their trivalent analogues due to the buffer effect of divalent TMs. This effect likely weakens as the ionic radius of the divalent TM increases. Therefore, divalent TMs such as $Co^{2+}$, $Ni^{2+}$ are more favorable than $Mn^{2+}$ and possibly $Fe^{2+}$, $Cu^{2+}$, $Zn^{2+}$. These guidelines are contrary to the common intuition that larger metal ions expand the lattice and are therefore favorable for Li transport. We further propose that the manipulation of SRO can also be relevant for the design and engineering of voltage profiles and oxygen activity through altering the local environments around Li, thereby enabling future optimization of this class of new cathode materials with unprecedentedly high capacities and compositional flexibility.

## Conclusions

Motivated by an experimental puzzle where two extremely similar compounds exhibit different electrochemical performance, we prove that cation short-range order (SRO) exists in long-range-disordered rocksalt cathodes. We have demonstrated how the SRO controls Li transport through altering local and macroscopic environments. More generally, we observe that electrostatics and ionic sizes strongly affect the SRO in disordered rocksalts of different chemistries. Our findings uncover an important direction for future engineering and optimization of disordered rocksalt cathodes to achieve higher capacities and energy densities.

## Methodology

**Synthesis**

To synthesize LMTO and LMZO, stoichiometric $Mn_2O_3$, $TiO_2$, $Zr(OH)_4$, and $Li_2CO_3$ (with 5% excess) were dispersed into ethanol and thoroughly mixed using a planetary ball mill (Retsch PM 200) at 300 rpm for 16 h. The mixture was then dried, pelletized, and calcinated at 1100 °C in an argon atmosphere for 2 h, followed by furnace cooling.

**Characterization**

The XRD patterns were collected using a Rigaku MiniFlex diffractometer equipped with a Cu source in the $2\vartheta$ range of 10–85°. Rietveld refinement was performed using the HighScore Plus software package. Elemental analysis was performed by Luvak Inc. using direct-current plasma emission spectroscopy (ASTM E 1097-12) for the quantitative identification of metal species. The oxygen contents were confirmed using the inert gas fusion method (ASTM E 1019-11). SEM images were obtained on a Zeiss Gemini Ultra-55 analytical field-emission scanning electron microscope. ED patterns were taken after a grain was oriented properly on JEM-2100F using selected area electron diffraction in TEM mode. Time-of-flight (TOF) neutron powder diffraction was performed at room temperature on the Nanoscale Ordered Materials Diffractometer (NOMAD) at the Spallation Neutron Source at Oak Ridge National Laboratory. The samples for the neutron experiment were synthesized using a 7-Li enriched $Li_2CO_3$ source. The PDF patterns were analyzed using the PDFGui software package[40].

**Electrochemistry**

To fabricate electrodes, the product powder was first shaker-milled (SPEX 8000) for 1 h in an argon atmosphere. The milled active material (70 wt%) was then manually mixed with Super C65 carbon black (Timcal, 20 wt%) in a mortar for 30 minutes, followed by mixing with polytetrafluoroethylene (Dupont, 10 wt%) in an Ar-filled glovebox. The mixture was then rolled into a thin film to be used as the cathode. A coin cell was assembled using 1 M $LiPF_6$ (in a volumetric 1:1 mixture of ethylene carbonate and dimethyl carbonate), glass microfiber filters (grade GF/F, Whatman), and Li metal foil as the electrolyte, separator, and anode, respectively. The coin cells were tested on an Arbin battery testing station. PITT measurements were performed on a Solartron Analytical 1470E Celltest System.

**DFT calculations**

All the DFT calculations were performed using the Vienna ab-Initio simulation package (VASP)[41, 42] with projector-augmented wave[43] pseudopotentials and the exchange-correlation functional by Perdew, Burke, and Ernzerhof.[44] To correct the DFT self-interaction error, the Hubbard-U correction[45] was employed for the transition-metal $d$ states where needed with values taken from Jain et al.[46] We employed $k$-point meshes with a reciprocal spacing of 25 $k$-points per Å$^{-1}$ for the Brillouin-zone integration and a plane wave basis set with an energy cutoff of 520 eV. All the DFT energies and atomic forces were converged to 0.001 meV/atom and 20 meV/Å, respectively. Input files for the DFT calculations were generated using the Python Materials Genomics[47] package.

**Cluster expansion**

Cluster expansion Hamiltonians[48] for each chemical space discussed here were constructed based on the energies of approximately 500 lattice configurations, where the energies were computed with DFT. Each cluster expansion relied on a basis set of pair interactions up to 7 Å, triplet interactions up to 4.1 Å, and quadruplet interactions up to 4.1 Å, with respect to a rocksalt primitive cell with lattice constant $a$ = 3 Å, on top of a baseline of formal-charge electrostatics and a fitted dielectric constant.[33] The effective cluster interactions and dielectric constant were obtained from a $L_1$-regularized linear regression fit, with the regularization parameter optimized by cross-validation.[49] The resulting fits yielded an out-of-sample root mean square error of less than 8 meV/atom. All canonical Monte Carlo simulations based on these Hamiltonians were run using the Metropolis-Hastings algorithm.

For the connectivity analysis, the definitions of 0-TM, 1-TM, and 2-TM channels of reference [4] were used. For each composition and temperature, the connectivity was averaged over 600 atomic configurations obtained from MC simulations as previously described. To ensure convergence of the 0-TM Li network size, 3×3×3 supercells of the MC configurations containing 12,960 cation sites were used.

**Simulation of ED patterns**

The LMZO and LMTO atomic configurations equilibrated with Monte-Carlo simulations at 1000°C as described above were used to simulate electron diffraction (ED) images. For this purpose, cubic sections with an edge length of 50 Å (>12000 atoms) were truncated from the periodic bulk structures defined by 960-atom unit cells. Electron diffraction patterns for the cubic [100] zone axis were computed for each cubic cell using the methods and potentials given by Kirkland,[50] using the potential calculation method described in reference [51]. The final diffraction images were calculated as an incoherent sum of all 500 simulated patterns, and these summed images were smoothed and scaled by amplitude (square root of intensity) to more clearly show the SRO features of the pattern.

## Acknowledgements

This work was supported by the Robert Bosch Corporation; Umicore Specialty Oxides and Chemicals; and the Assistant Secretary for Energy Efficiency and Renewable Energy, Vehicle Technologies Office of the U.S. Department of Energy under Contract No. DE-AC02-05CH11231 under the Advanced Battery Materials Research (BMR) Program. The research conducted at the NOMAD Beamline at ORNL's Spallation Neutron Source was sponsored by the Scientific User Facilities Division, Office of Basic Sciences, U.S. Department of Energy. Work at the Molecular Foundry was supported by the Office of Science, Office of Basic Energy Sciences, of the U.S. Department of Energy under Contract No. DE-AC02-05CH11231.The computational analysis was performed using computational resources sponsored by the Department of Energy's Office of Energy Efficiency and Renewable Energy and located at the National Renewable Energy Laboratory, as well computational resources provided by Extreme Science and Engineering Discovery Environment (XSEDE), which was supported by National Science Foundation grant number ACI-1053575. The authors would like to thank Jinhyuk Lee, Jue Liu, Yuanpeng Zhang, and Penghao Xiao for helpful discussion.

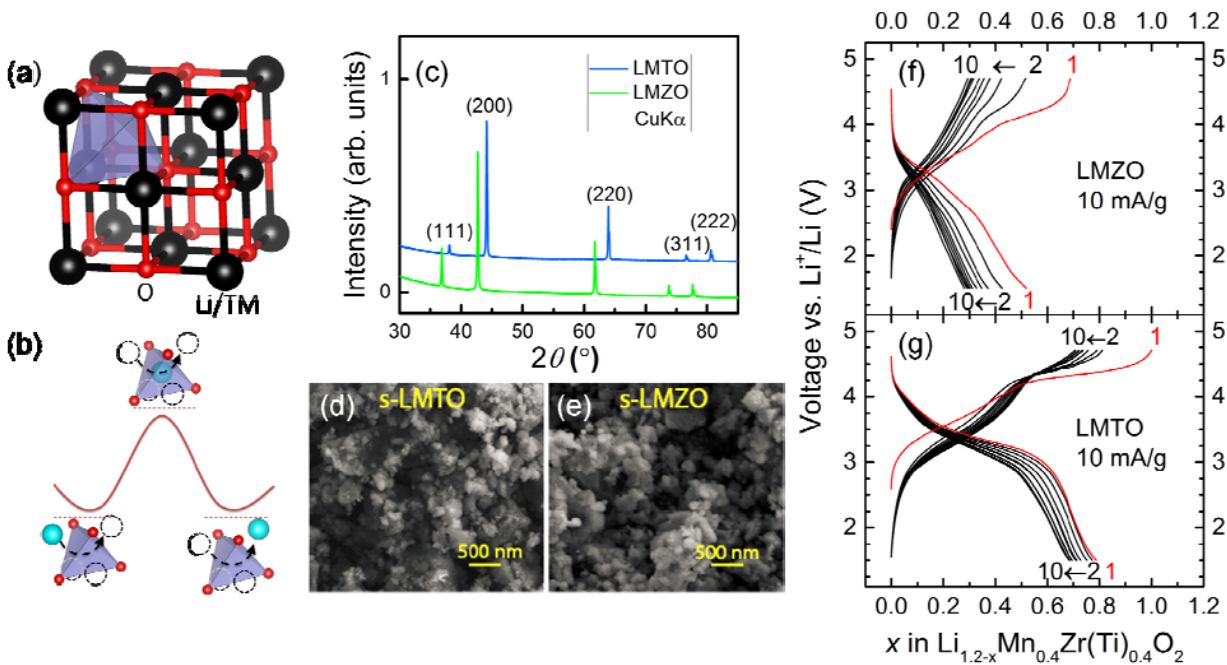

**Figure 1.** (a) Crystal structure of an ideal cation-disordered rocksalt-type lithium metal oxide. The black spheres represent metal cations (including lithium and TMs), and the red spheres represent oxygen anions. Both cations and anions are in octahedral coordination. The highlighted blue tetrahedral site represents a typical migration pathway for Li diffusion. (b) Schematic energy landscape of Li migration from its octahedral coordination through a tetrahedral vacancy into another octahedron. The energy barrier depends on the local environment and size of the tetrahedron. The migrating Li ion is highlighted in cyan. (c) XRD patterns of LMZO and LMTO indexed according to the rocksalt structure. The low-angle shift in the pattern of LMZO compared with that of LMTO indicates the larger lattice parameter of LMZO. (d–e) SEM images of shaker-milled LMTO (s-LMTO) and LMZO (s-LMZO) with similar particle sizes of ~100 nm. (f–g) Voltage profiles of LMZO and LMTO between 1.5 and 4.7 V for the first 10 cycles at room temperature.

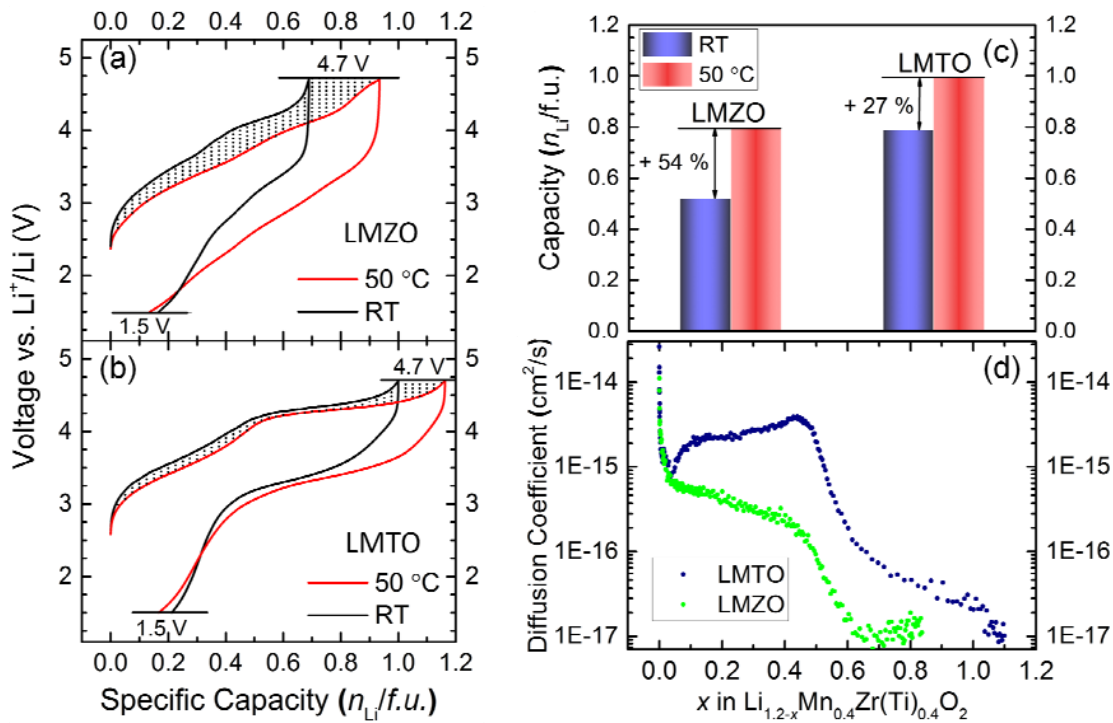

**Figure 2.** Comparison of high-temperature and room-temperature galvanostatic cycling of LMTO and LMZO, and PITT measurements. First-cycle galvanostatic voltage profiles of (a) LMZO and (b) LMTO at 50 °C and room temperature. The shaded area represents the difference between the high-temperature and room-temperature charge profiles. (c) First-cycle reversible capacities of LMZO and LMTO at high temperature and room temperature. (d) Li chemical diffusion coefficients of LMZO (green) and LMTO (blue) obtained from fitting the room-temperature PITT data at various Li contents.

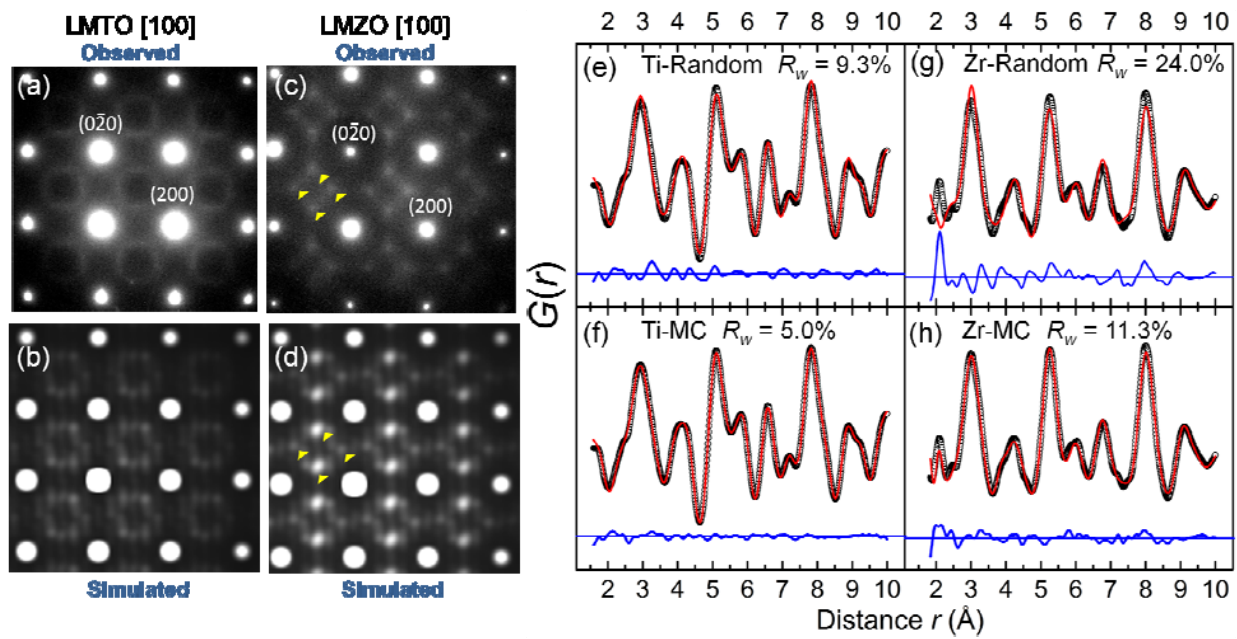

**Figure 3.** Experimental observation and computational simulation of short-range order (SRO) in LMTO and LMZO. ED patterns of LMTO (a) and LMZO (c) along the zone axis [100]. The round spots are indexed to the F$m$-3$m$ space group, while the diffuse scattering patterns nearby are attributed to SRO. Several intensity maxima in the diffuse scattering patterns are highlighted with yellow arrows in LMZO. Simulation of ED patterns for LMTO (b) and LMZO (d) along the same zone axis shows good agreement with experimental observation. Refinement of NPDF data of LMTO (e, f) and LMZO (g, h) using the random model (e, g) and MC-equilibrated structural models (f, h). The experimental data are plotted as black open circles. The calculated values are plotted as solid red lines. The difference between observation and calculation is plotted as solid blue lines.

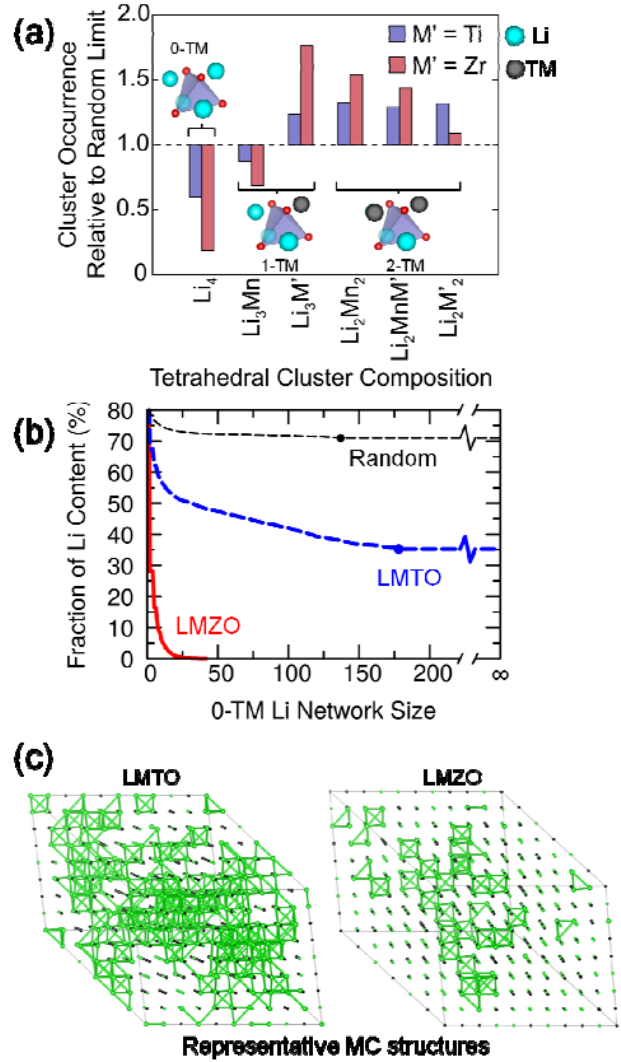

**Figure 4.** Local cation cluster and macroscopic Li connectivity analysis based on MC-derived structures for LMTO and LMZO at 1000°C. Each MC structure contains 480 cation sites, of which 288 are decorated with Li ions. (a) Occurrence of various tetrahedral clusters (0-TM, 1-TM, 2-TM) in LMTO (blue) and LMZO (red) as compared to the random limit. (b) Connectivity plots of LMZO and LMTO showing the fraction of Li content in networks of at least a certain number of Li sites. A Li network is defined as all the Li sites that are interconnected through 0-TM channels. Each plot is averaged over 600 sampled MC structures. The result for a random cation distribution with the same Li to TM ratio is also plotted as a reference. The plots are truncated

at the percolating Li contents (marked by dots) and extended to infinity for LMTO and the random case. LMZO is not percolating. (c) Representative MC structures for LMTO and LMZO. Li ions are labeled with green spheres and 0-TM connected Li sites are bridged with green bonds.

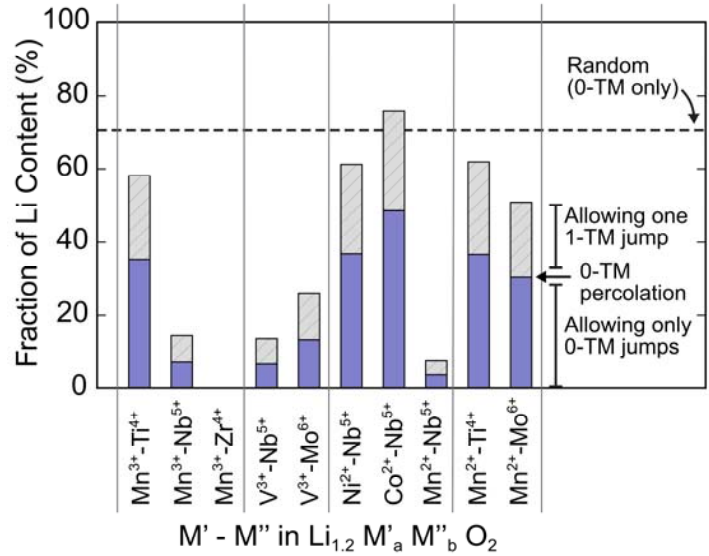

**Figure 5.** Fraction of Li content made accessible by the percolating network, considering only 0-TM jumps (blue) or allowing any given Li to make a single 1-TM jump before reaching the 0-TM percolating network (grey), in a specific $Li_{1.2}M'_aM''_bO_2$. The dotted line marks the fraction of Li accessible within the 0-TM percolating network in the random structure limit. The stoichiometry of each $Li_{1.2}M'_aM''_bO_2$ compound is constructed such that charge neutrality is retained. The various combinations of TM species M'–M'' are indicated along the $x$-axis.

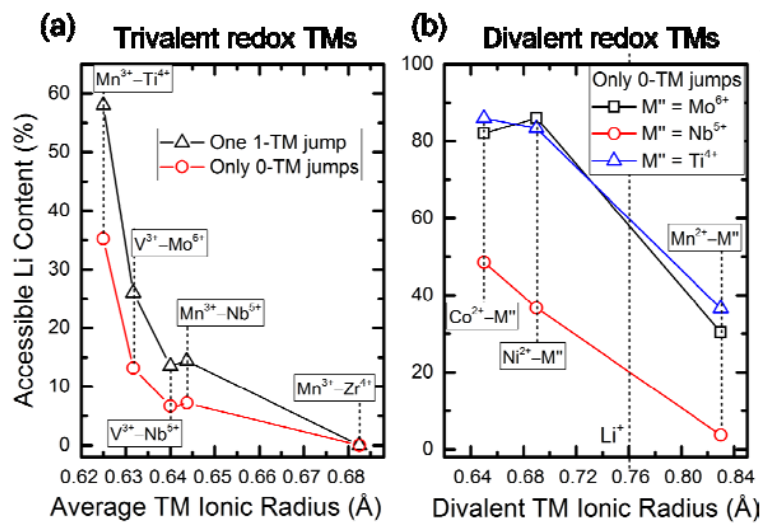

**Figure 6.** Correlation between accessible Li contents and ionic radii in various DRXs. (a) Accessible Li content as a function of the average TM ionic radius in DRXs composed of trivalent redox-active TMs, by allowing only 0-TM jumps (red) or allowing any given Li to make a single 1-TM jump before reaching the 0-TM percolating network (black). (b) Accessible Li content as a function of the divalent TM ionic radius in DRXs with divalent redox-active TMs and various stabilizers by allowing only 0-TM jumps. The ionic radius of Li$^+$ is marked at 0.76 Å.

# Supplementary Information

# Hidden structural order controls Li-ion transport

# in cation-disordered oxides for rechargeable lithium batteries

Huiwen Ji, Alexander Urban, Daniil A. Kitchaev, Deok-Hwang Kwon, Nongnuch Artrith, Colin Ophus,

Wenxuan Huang, Zijian Cai, Tan Shi, Jae Chul Kim, Gerbrand Ceder[*]

**Table S1.** Results of elemental analysis by direct-current plasma emission spectroscopy

| Materials | Target<br><br>Li: Mn: Ti: Zr | Measured<br><br>Li: Mn: Ti: Zr |
|---|---|---|
| $Li_{1.2}Mn_{0.4}Ti_{0.4}O_2$ | 1.2: 0.4: 0.4: 0 | 1.227: 0.373: 0.399: 0 |
| $Li_{1.2}Mn_{0.4}Zr_{0.4}O_2$ | 1.2: 0.4: 0: 0.4 | 1.218: 0.412: 0: 0.370 |

The compositions of the products were verified to be nearly identical to the target compositions using elemental analysis.

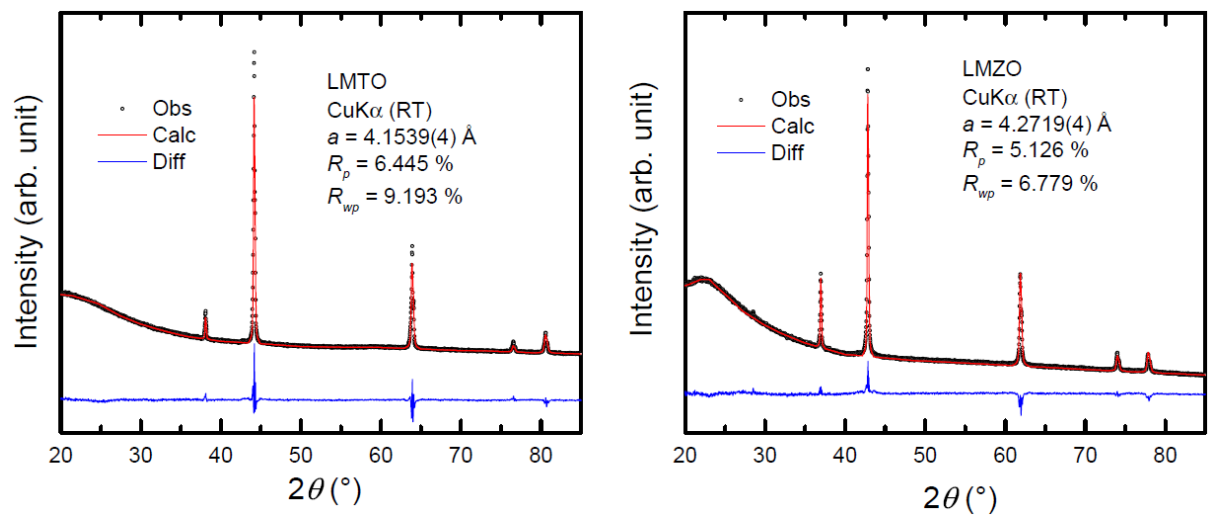

**Figure S1.** Rietveld refinement of XRD patterns of $Li_{1.2}Mn_{0.4}Ti_{0.4}O_2$ (LMTO, left) and $Li_{1.2}Mn_{0.4}Zr_{0.4}O_2$ (LMZO, right). A rocksalt structural model (space group $Fm$-$3m$) was used for the refinement. The experimental data are plotted as black open circles. The calculated values based on structural models are plotted as solid red lines. The difference between observation and calculation is plotted as solid blue lines. Only the instrumental zero shift, scale factor, lattice constants, and peak-profile parameters $U$, $V$, $W$ were refined. The isotropic thermal parameter $B_{iso}$ was fixed at a typical value of 0.5 Å$^2$. The site occupancies were set to those of the target compositions and were not refined.

**Ex-situ XRD of LMZO**

Because LMZO is a new compound first reported in this work, we performed ex-situ XRD on the cycled products to determine its stability over cycling.

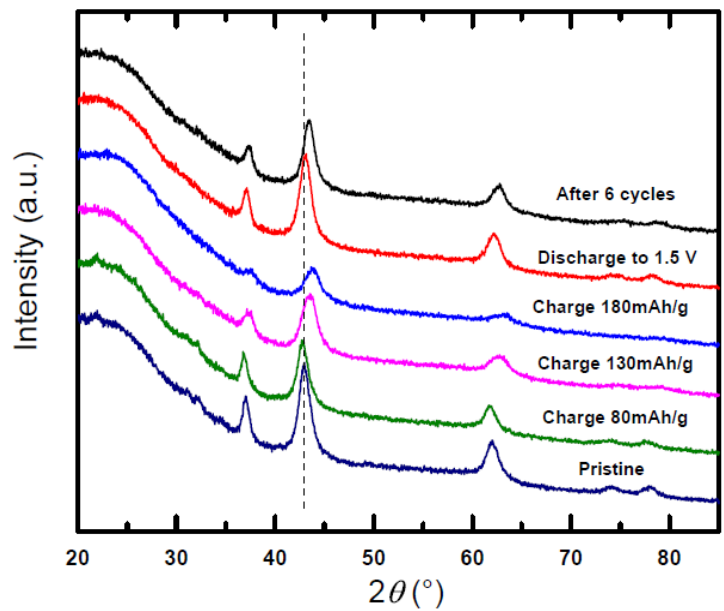

**Figure S2.** Ex-situ XRD patterns of LMZO during the first cycle and after 6 cycles. The cathode films were galvanostatically cycled at 10 mA/g at room temperature. The charging capacity 180 mAh/g roughly corresponds to the extractable capacity with an upper cutoff voltage of 4.7 V. The patterns show a reversible change in lattice parameters, with X-ray peaks recovered to the initial positions of the pristine film. After 6 cycles, the cation-disordered rocksalt structure was retained although the lattice parameter slightly shrank as compared to the pristine sample, indicated by the right-shifted X-ray peaks. The lattice contraction may be associated with lattice densification, which is a common reason that leads to capacity fading in cation-disordered Li-TM oxide cathode materials[1].

**Computed voltage profiles of LMTO and LMZO**

The voltage curves were computed by first enumerating a large number of possible configurations for a given cathode material at various Li levels. After collecting the ground state configurations, the voltage profiles were then computed using the battery builder function from pymatgen.[2]

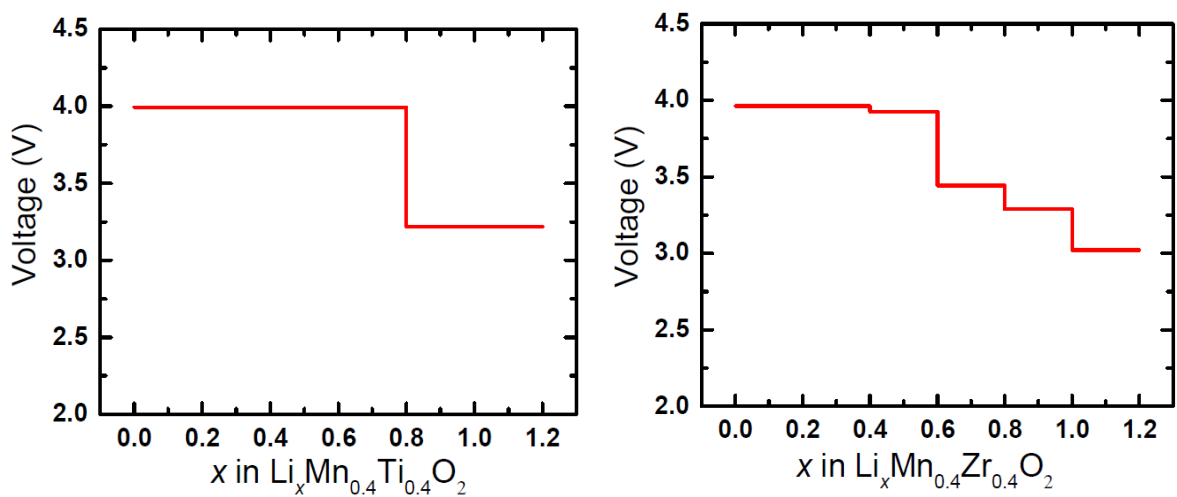

**Figure S3.** Computed voltage profiles of LMTO (left) and LMZO (right).

The computed voltage curves indicate that both materials are predicted to operate in a similar voltage window despite their slightly different voltage profiles. Therefore, the performance of LMZO is unlikely limited by thermodynamics. The open-circuit voltage of LMZO is calculated to be ~ 0.2 V lower than that of LMTO, consistent with the observation from electrochemical tests.

**Li chemical diffusivity analysis**

The PITT tests were performed on LMTO and LMZO. Both materials were charged from the open-circuit voltages to 4.7 V with a 0.01-V incremental step interval. At each step, the voltage was held constant for 1 h, and the corresponding current–time response was recorded. To determine the Li chemical diffusivity, an equation developed by Wen, Boukamp, and Huggins[3] based on Fick's second law was used. The equation was applied in the short-time approximation as follows:

$$I(t) = \frac{QD^{1/2}}{L\pi^{1/2}} \frac{1}{\sqrt{t}} \qquad t \ll \frac{L^2}{D}$$

In the equation, $I$ is the current; $D$ is the Li chemical diffusivity; and $L$ represents the diffusion length, which in this case is half of the average particle size, 50 nm. Within the short-time region (typically within 400 s), a plot of $I$ as a function of $t^{-1/2}$ should give a linear response. The Li diffusivity $D$ can be extracted from the slope of this linear region. Theoretically, both the short-time and long-time approximations should give the same results. However, the long-time approximation condition was never fulfilled within the limited time frame of this experiment and with the slow kinetics of the materials.

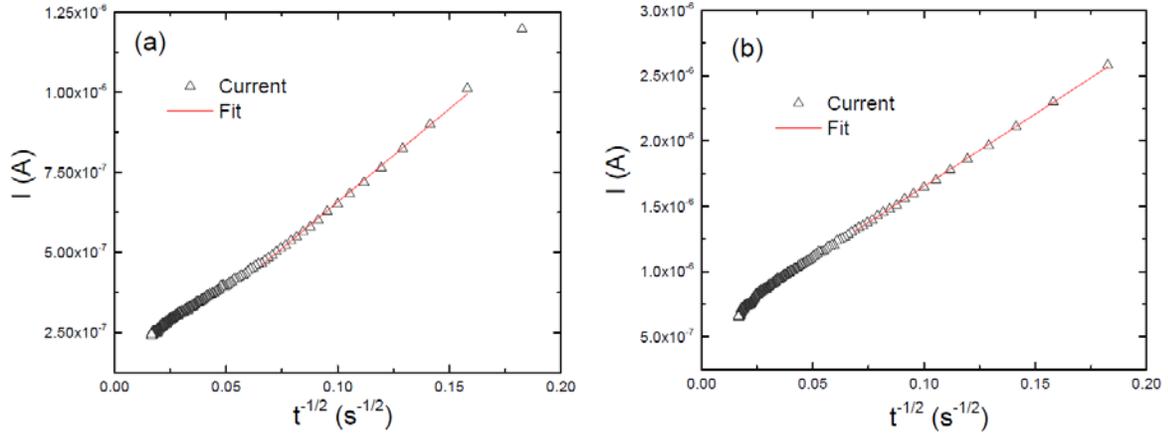

**Figure S4.** Examples of fitting PITT data for LMTO (a) and LMZO (b). The current–time response was recorded at 2.803 V for LMTO and at 2.604 V for LMZO and is plotted with black triangles. The linear fittings are plotted with solid red lines.

The slope from fitting $I$ vs. $t^{-1/2}$ for LMTO is $5.810 \times 10^{-6}$ C·s$^{-1/2}$. The accumulated charge for this voltage step is 0.001162 C, and the diffusion length $L$ is 50 nm. Therefore, the Li chemical diffusivity is $1.963 \times 10^{-15}$ cm$^2$/s for this step. Likewise, the slope from fitting $I$ vs. $t^{-1/2}$ for LMZO is $1.094 \times 10^{-5}$ C·s$^{-1/2}$. The accumulated charge is 0.003174 C. The resulting Li diffusivity is $9.324 \times 10^{-16}$ cm$^2$/s.

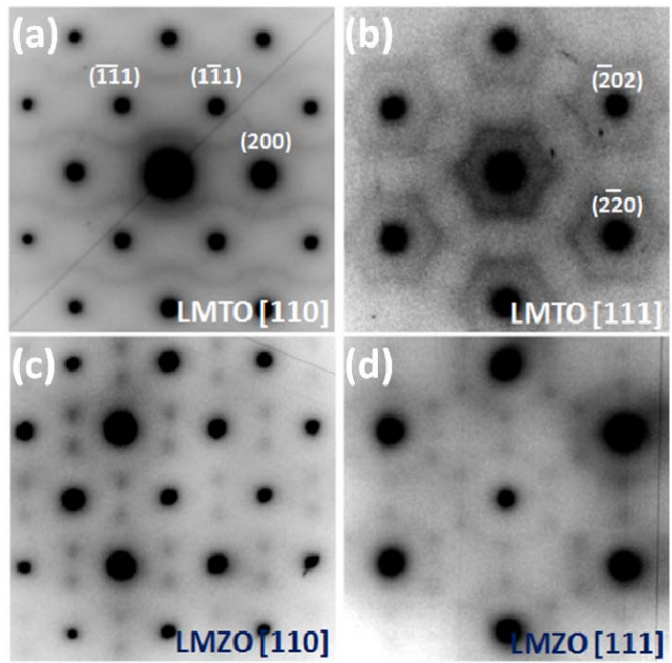

**Figure S5.** ED patterns of LMTO (a, b) and LMZO (c, d) along the zone axes [110] and [111]. Along each zone axis, the images of LMTO and LMZO are aligned towards the same orientation. The round Bragg spots observed are indexed to the F$m$-3$m$ space group, while the diffuse scattering patterns are attributed to SRO. Consistent with the observation along [100] in Figure 3a and 3c, the diffuse scattering patterns are completely different for the two materials, suggesting significant difference in their SRO. In addition, the diffuse scattering patterns of LMZO show intensity maxima that do not exist in the patterns of LMTO, suggesting more pronounced SRO in LMZO.

**Technical details of NPDF refinement**

**Table S2.** Refinement results of NPDF data for LMTO and LMZO in a short-$r$ range between the occurrence of the shortest M-O bond (1.6 Å for LMTO and 1.8 Å for LMZO) and 10 Å using the random structural model

|  | LMTO (short-$r$ range) | LMZO (short-$r$ range) |
|---|---|---|
| Space Group | $Fm$-$3m$ | $Fm$-$3m$ |
| $a$ (Å) | 4.1674 | 4.2837 |
| $\delta_1$* | 0.26213 | 1.6 |
| $U_{iso}$ (Å$^2$) | 0.019497 | 0.03134 |
| $R_W$ | 9.3% | 24.0% |

**Table S3.** Refinement results of NPDF data for LMTO and LMZO in a long-$r$ range between 40 Å and 50 Å using the random structural model

|  | LMTO (long-$r$ range) | LMZO (long-$r$ range) |
|---|---|---|
| Space Group | $Fm$-$3m$ | $Fm$-$3m$ |
| $a$ (Å) | 4.1582 | 4.2740 |
| $\delta_1$* | 0.26213 | 1.6 |
| $U_{iso}$ (Å$^2$) | 0.014208 | 0.0186 |
| $R_W$ | 12.5% | 12.2% |

*$\delta_1$ is a high-temperature vibrational correlation parameter and modifies peak widths at short-$r$ values. Therefore, $\delta_1$ was only refined within the short-$r$ range. The obtained value was then used for the long-$r$ range without further refinement.

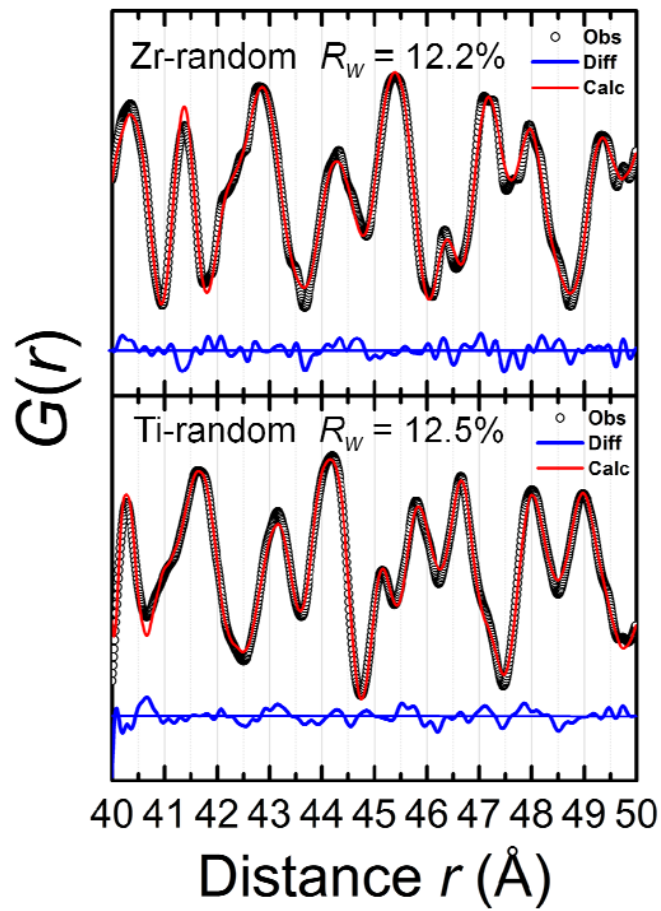

**Figure S6.** Refinement of NPDF data of LMZO (top) and LMTO (bottom) using the random structural model in a long-*r* range. The experimental data are plotted as black open circles. The calculated values based on structural models are plotted as solid red lines. The difference between observation and calculation is plotted as solid blue lines.

**Refinement of NPDF data within the short-*r* range using MC-equilibrated structures**

For each composition, 10 atomic configurations (each with 60 cations) were first obtained from MC equilibration at 1273 K. The atomic coordinates and lattice parameters of all the structures were relaxed using DFT to capture local variations in bond lengths. Each MC structure was then individually fit to the experimental pattern of the corresponding composition. A typical refinement for LMZO using one MC structure is shown in Figure S7. The MC structure does not have any symmetry and is in Space Group $P1$. The lattice constants $a$, $b$, $c$, angles $\alpha$, $\beta$, $\gamma$, vibrational correlation parameter $\delta_1$, and isotropic thermal parameter $U_{iso}$ were refined, and the results are presented in Table S4. After individual fitting, we numerically averaged the calculated profiles based on the 10 configurations and produced a one-dimensional PDF $G(r)$, which was eventually compared with the experimental $G(r)$, as shown in Figure 3f and 3h in the main manuscript.

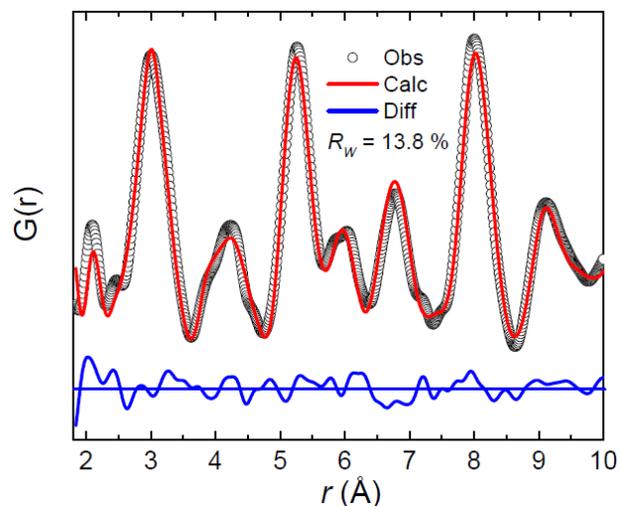

**Figure S7.** An example of NPDF refinement for LMZO using one MC structure within the short-$r$ range from 1.8 Å to 10 Å. The experimental data are plotted as black open circles. The calculation based on the MC structural model is plotted as a solid red line. The difference between observation and calculation is plotted as a solid blue line.

**Table S4.** Refinement results of NPDF data for LMZO within the short-$r$ range (from 1.8 Å to 10 Å) using one MC structure

|  | LMZO (short-$r$ range) |
| --- | --- |
| Space Group | $P1$ |
| $a$ (Å) | 9.11118 |
| $b$ (Å) | 10.8747 |
| $c$ (Å) | 13.3222 |
| $\alpha$ | 68.2145 |
| $\beta$ | 81.5379 |
| $\gamma$ | 74.7896 |
| $\delta_1$ | 1.54324 |
| $U_{iso}$ (Å$^2$) | 0.009486 |
| $R_W$ | 13.8% |

**Table S5.** Occurrence of various tetrahedral clusters in LMTO and LMZO[*]

| Cluster composition | $Li_{1.2}Mn_{0.4}Ti_{0.4}O_2$ 1000 °C | $Li_{1.2}Mn_{0.4}Zr_{0.4}O_2$ 1000 °C | $Li_{1.2}Mn_{0.4}M'_{0.4}O_2$, Random limit |
|---|---|---|---|
| **Tetrahedral clusters** | | | |
| $Li_4$ | 0.076 | 0.024 | 0.127 |
| $Li_3M$ | 0.367 | 0.426 | 0.348 |
| $Li_3Mn$ | 0.152 | 0.119 | 0.174 |
| $Li_3M'$ | 0.215 | 0.307 | 0.174 |
| $Li_2M_2$ | 0.454 | 0.479 | 0.348 |
| $Li_2Mn_2$ | 0.115 | 0.135 | 0.087 |
| $Li_2MnM'$ | 0.224 | 0.250 | 0.174 |
| $Li_2M'_2$ | 0.115 | 0.095 | 0.087 |

[*] Only Li-rich tetrahedral clusters relevant to Li migration (0-TM, 1-TM, 2-TM) are listed.

**Details about connectivity analysis**

The connectivity analysis is performed on MC structures equilibrated at 1000°C. Each MC structure contains 480 cation sites, of which 288 are decorated with Li ions. The connectivity function is defined as following:

$$P(n) = Pr\left\{\sum_{j=1}^{N} I(u_j; n) = 1\right\} /N \times 100$$

where $I(u_j; n)$ is an indicator to record whether Li ion $u_j$ is in a Li network of at least $n$ Li sites. If the Li ion $u_j$ is in a Li network of at least $n$ Li sites, then $I(u_j; n) = 1$; otherwise $I(u_j; n) = 0$. $N$ is the total number of Li sites in a MC structure. The resulting $P(n)$'s are averaged over the 600 sampled MC structures for each composition to obtain the connectivity plots shown in Figure 4(b).

**Nearest-neighbor pair (NNP) parameters for effective interaction between cation species**

We present the nearest-neighbor pair (NNP) parameters for six representative compositions ($Li_{1.2}Mn_{0.4}Ti_{0.4}O_2$, $Li_{1.2}Mn_{0.4}Zr_{0.4}O_2$, $Li_{1.2}Mn_{0.6}Nb_{0.2}O_2$, $Li_{1.2}Ni_{0.4}Nb_{0.4}O_2$, $Li_{1.2}Co_{0.4}Nb_{0.4}O_2$, and $Li_{1.2}Mn_{0.4}Nb_{0.4}O_2$) in Figure S8. By comparing the occurrence of a cation pair in the first neighboring shell in MC structures to that in a random structure, net attraction (negative NNPs) or repulsion (positive NNPs) can be extracted between the two cation species.

For each composition, 600 atomic configurations sampled from MC simulations as previously described were analyzed and averaged to obtain the reported nearest-neighbor statistics. The NNP parameter is defined as following:

$$NNP(A-B) = 1 - \frac{n_{random}(B)}{n_{MC}(B)},$$

where A and B are two types of cations; $n_{MC}(B)$ is the number of cations of type B within the first neighboring shell around cation A averaged over 600 MC structures; and $n_{random}(B)$ is the number of cations of type B within the first neighboring shell surrounding cation A based on a random model. If NNP(A–B) < 0, the effective interaction between A and B ions is attractive; if NNP(A–B) > 0, the interaction is repulsive; and if NNP(A–B) = 0, the interaction vanishes. For example, if we consider the pair of $Li^+$ and $Ti^{4+}$ in LMTO, the statistical number of $Ti^{4+}$ ions within the first neighboring shell of $Li^+$ is $12 \times 0.2 = 2.4$, whereas the average number of $Ti^{4+}$ ions counted from the MC structures is 2.769. Therefore, NNP($Li^+$–$Ti^{4+}$) is −0.154, indicating net attractive interaction between the two species.

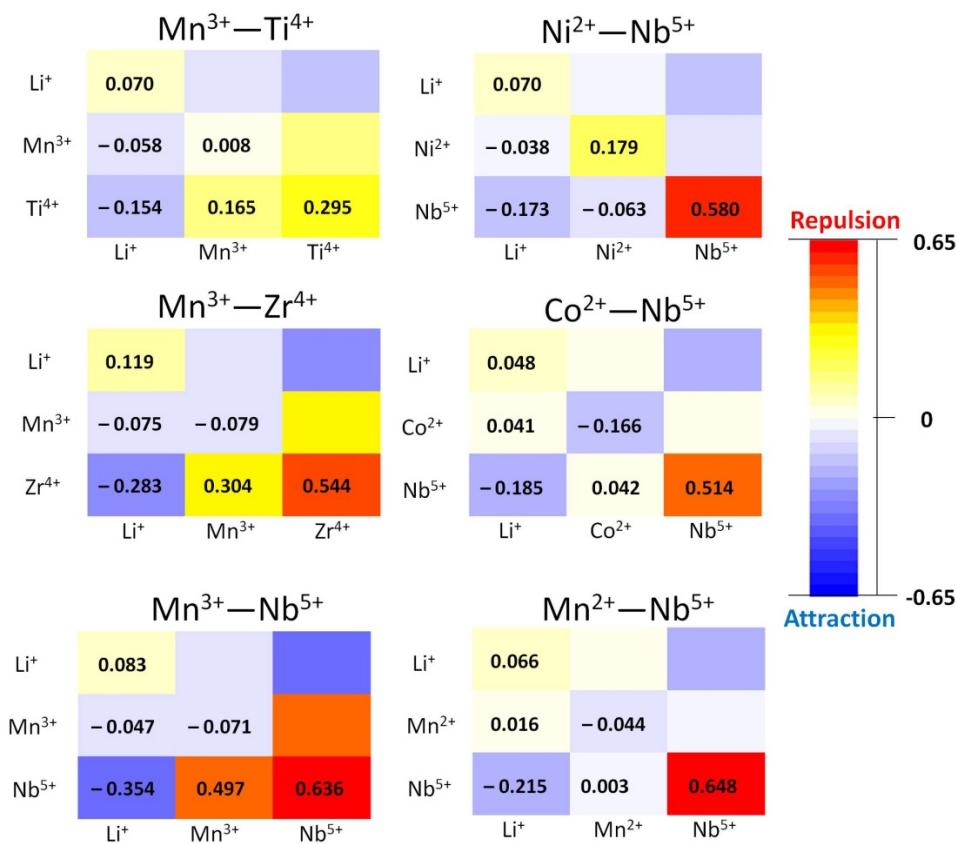

**Figure S8.** Diagrams of Nearest-neighbor pair parameters for various $Li_{1.2}M'_aM''_bO_2$ compounds. The color of a block visualizes the attraction (blue) or repulsion (red) between two metal species.

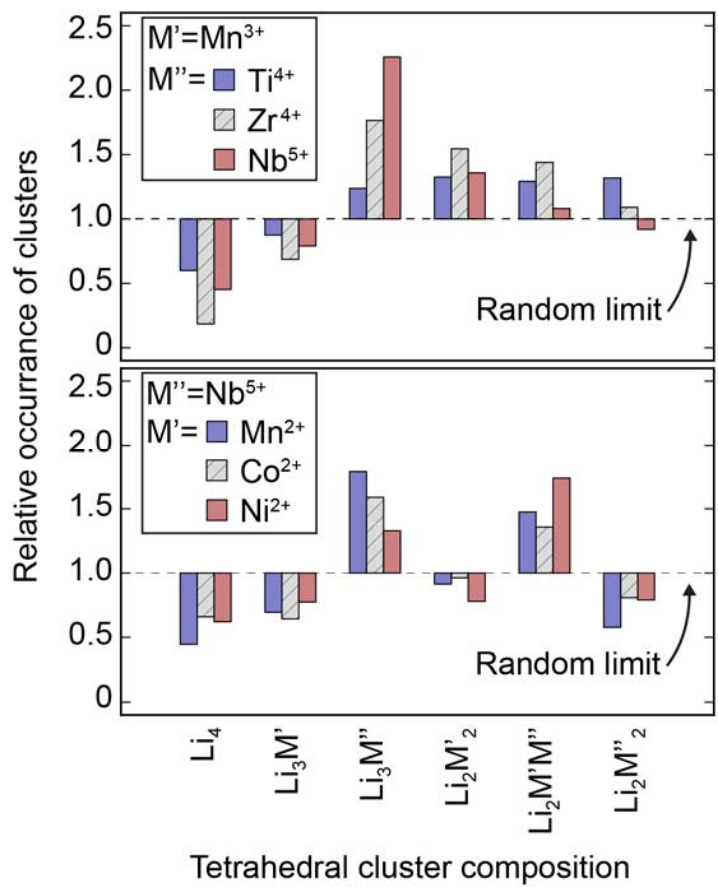

Figure S9. Occurrence of tetrahedral clusters in various $Li_{1.2}M'_aM''_bO_2$ compounds relative to the random limit. The stoichiometry of the each $Li_{1.2}M'_aM''_bO_2$ compound is constructed such that charge neutrality is retained.

**Supplementary References**